\documentclass[10pt,twocolumn,english,aps,prd,nofootinbib,showkeys,preprintnumbers,showpacs]{revtex4}
\usepackage[utf8]{inputenc}
\usepackage{graphicx}
\usepackage{amsmath}
\usepackage{amssymb}
\usepackage{amsfonts}
\usepackage{lscape}
\usepackage{hyperref}
\usepackage{url}
\usepackage{epsfig}
\usepackage{color}
\usepackage{bm}
\usepackage{tabularx}
\newcommand{\be}{\begin{equation}}
\newcommand{\ee}{\end{equation}}
\newcommand{\ba}{\begin{eqnarray}}
\newcommand{\ea}{\end{eqnarray}}

\newcommand{\rr}{\mathrm}

\newcommand{\OPBH}{\Omega_{\rm PBH}}
\newcommand{\OCDM}{\Omega_{\rm CDM}}
\newcommand{\Msun}{M_\odot}
\newcommand{\Mtot}{M_{\rm cl}}
\newcommand{\Ncl}{N_{\rm cl}}

\newcommand{\Mpbh}{M_{\rm PBH}}
\newcommand{\fpbh}{f_{\rm PBH}}

\makeatother

\begin{document}

\title{Constraints from microlensing experiments on clustered primordial black holes}

\author{Juan Garc\'ia-Bellido$^{(a,b)}$ and S\'ebastien Clesse$^{(c,d,e)}$}
\email{juan.garciabellido@cern.ch, \\ sebastien.clesse@unamur.be}
\affiliation{$^{(a)}$Instituto de F\'isica Te\'orica UAM-CSIC, Universidad Auton\'oma de Madrid,
Cantoblanco, 28049 Madrid, Spain\\
$^{(b)}$CERN, Theoretical Physics Department, 1211 Geneva, Switzerland\\
$^{(c)}$ Centre for Cosmology, Particle Physics and Phenomenology, Institute of Mathematics and Physics, Louvain University, 2 Chemin du Cyclotron, 1348 Louvain-la-Neuve, Belgium \\
$^{(d)} $Namur Center of Complex Systems (naXys), Department of Mathematics, University of Namur, Rempart de la Vierge 8, 5000 Namur, Belgium \\
$^{(e)}$Institute for Theoretical Particle Physics and Cosmology (TTK), RWTH Aachen University, D-52056 Aachen, Germany}

\preprint{CERN-TH-2017-206, IFT-UAM/CSIC-17-094}

\date{\today}

\begin{abstract}
It has recently been proposed that massive primordial black holes (PBH) could constitute all of the dark matter, providing a novel scenario of structure formation, with early reionization and a rapid growth of the massive black holes at the center of galaxies and dark matter halos. The scenario arises from broad peaks in the primordial power spectrum that give both a spatially clustered and an extended mass distribution of PBH. The constraints from the observed microlensing events on the extended mass function have already been addressed. Here we study the impact of spatial clustering on the microlensing constraints. We find that the bounds can be relaxed significantly for relatively broad mass distributions if the number of primordial black holes within each cluster is typically above one hundred. On the other hand, even if they arise from individual black holes within the cluster, the bounds from CMB anisotropies are less stringent due to the enhanced black hole velocity in such dense clusters. This way, the window between a few and ten solar masses has opened up for PBH to comprise the totality of the dark matter.
\end{abstract}

\maketitle

\section{Introduction}

Since the detection of gravitational waves from the merging of five massive black hole binaries by the Advanced LIGO/VIRGO interferometers~\cite{Abbott:2016blz,Abbott:2016nmj,TheLIGOScientific:2016pea,Abbott:2017vtc,Abbott:2017oio}, see Table 1, a lot of attention has been given to the possibility that these massive black holes could actually be of primordial origin and that they could constitute all of the dark matter~\cite{Bird:2016dcv,Clesse:2016vqa,Sasaki:2016jop}. Scenarios of PBH production from large peaks in the matter power spectrum that could provide the totality of the dark matter date back several decades~\cite{GarciaBellido:1996qt}, and more recently it has been suggested that the distribution of PBH is a lognormal in mass and that PBH are spatially clustered~\cite{Clesse:2015wea}, as one would expect from a broad peak in the primordial power spectrum~\cite{Chisholm:2005vm}.

The scenario we are considering here~\cite{Garcia-Bellido:2017fdg} is that of cold dark matter comprised of compact clusters of several hundreds to thousands of PBH in a small volume, with a very massive black hole at the centre that has grown due to merging from dynamical friction, and a swarm of smaller but still massive black holes orbiting closely around it, sometimes colliding and merging~\cite{Clesse:2016ajp}, others simply scattering off each other, emitting gravitational waves in the process~\cite{Garcia-Bellido:2017qal}. Such compact clusters behave like collisionless cold dark matter ``particles" falling in the potential wells left by curvature fluctuations generated during inflation, giving rise to the present large scale structures observed in deep galaxy surveys like SDSS and DES~\cite{Abbott:2017wau}.

The most stringent constraints on PBH in the range of a few solar masses come from microlensing experiments~\cite{Carr:2016drx}, like MACHOS, EROS~\cite{Alcock:1998fx,Tisserand:2006zx} and OGLE~\cite{Wyrzykowski:2015ppa}, and from dwarf spheroidals, like Eridanus II~\cite{Brandt:2016aco,Green:2016xgy,Li:2016utv}, and much less strongly from CMB anisotropies~\cite{Poulin:2017bwe,Ali-Haimoud:2016mbv}. Interestingly, some massive compact objects have been detected by microlensing experiments~\cite{Wyrzykowski:2015ppa,Mediavilla:2017bok}, as well as by their radio emission in dense molecular clouds~\cite{Oka:2017}.   Other clues for PBH Dark Matter include the spatial correlation in the cosmic infrared background and soft x-ray backgrounds~\cite{Kashlinsky:2016sdv,Cappelluti:2017nww} and the detection of a huge population of super-massive black holes at high redshifts in the Chandra deep field~\cite{Luo:2016ojb}.  PBH could also provide natural mechanisms to resolve the small scale crisis of large scale structure~\cite{CGB}. In the future, more results will come from GAIA astrometry~\cite{GAIA} as well as from a follow up of the Fermi-LAT point source catalog~\cite{Acero:2015hja}.

The wide mass and low spin distributions of the observed LIGO Black Hole Binaries (BHB)~\cite{Abbott:2017vtc} come naturally from early universe models of PBH formation from large peaks in the matter power spectrum, arising both in single (e.g.~\cite{Garcia-Bellido:2017mdw,Ezquiaga:2017fvi}) and multi-field (e.g.~\cite{Garcia-Bellido:2016dkw,Garcia-Bellido:2017aan}) models of inflation. These models predict in general a broad peak in the power spectrum of fluctuations, which leads to a wide mass distribution as well as to significant clustering of those primordial black holes. 

The effect of wide mass distributions on the general PBH constraints has already been addressed in Ref.~\cite{Carr:2017jsz,Bellomo:2017zsr}. In this paper we leave out most of the discussion on future prospects of detectability of PBH~\cite{CGB}, and concentrate on the modification of the PBH constraints due to their clustering.  These can be very different if BH are uniformly distributed or, on the contrary, if they are hierarchically clustered, with the more massive black hole at the center of the cluster and the less massive ones orbiting around them. In the latter case, the spatial distribution is more like a complicated and the probability of one given PBH cluster being in the line of sight of a particular star, in say the Large Magellanic Cloud, is significantly reduced.

In the next section we will study the lognormal distribution of black hole masses reconstructed from the {\em known} AdvLIGO events and will discuss the effect that such a broad distribution has on synthetic microlensing contraints. Next we will study the effect of clustering on the microlensing and CMB constraints and will apply the formalism to the present PBH constraints. Finally, we will present our conclusions.

\section{The mass distribution of PBH}

\begin{figure}[!ht]
\centering
\includegraphics[width = 0.48\textwidth]{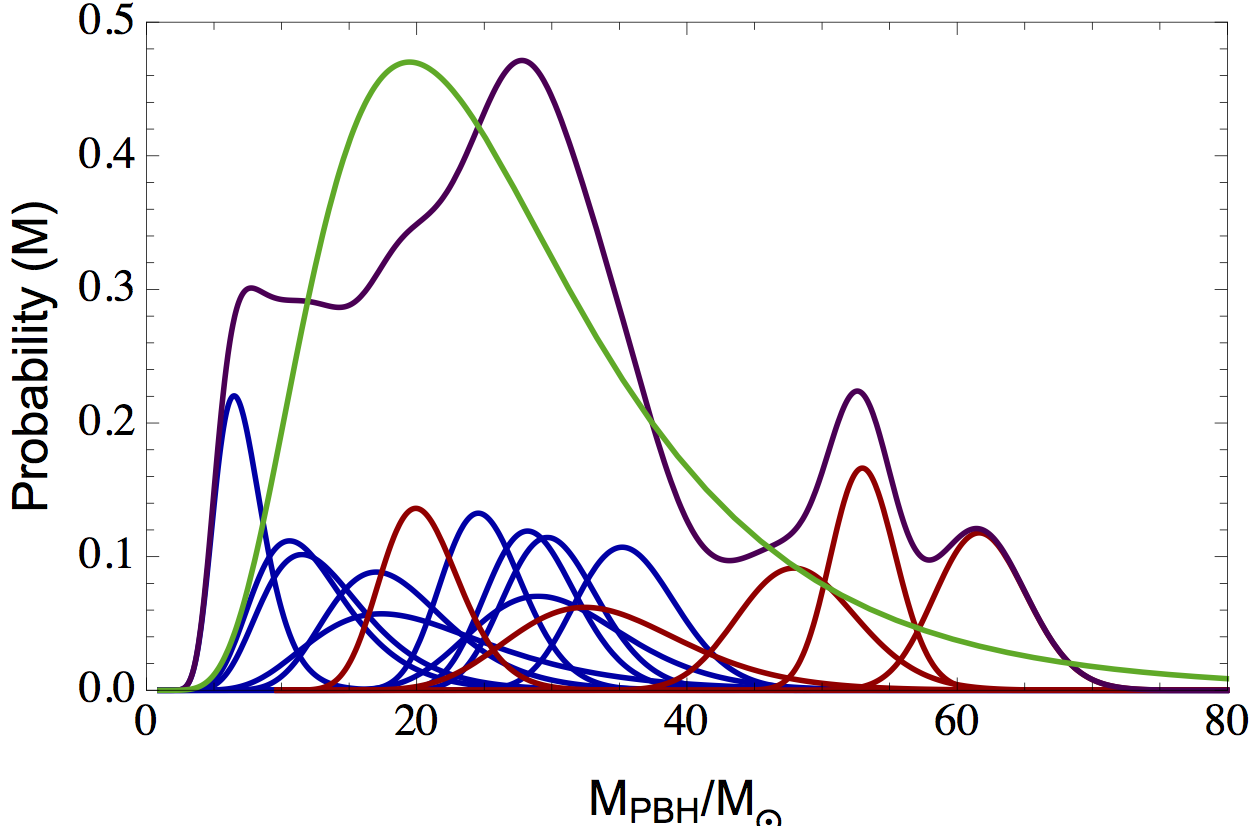}
\caption{The individual mass distributions of the observed LIGO black holes before (blue curves) and after (red curves) merging. The sum of all individual black hole mass distributions, including the final merged ones (purple). Taking all of them into account, one obtains a weighted sum of the Adv\-LIGO black hole mass distribution, which is a lognormal with $\mu=25\,\Msun$ and $\sigma=0.5$ (green curve). }
\vspace*{-1mm}
\label{fig:LIGO}
\end{figure}

The five BHB mergers detected by Advanced LIGO and recently also by the Virgo Collaboration are distributed in masses in the range from 8 to 40 $\Msun$. The final black holes after merger have masses between 20 and 70 $\Msun$.
All the masses can be seen in Table~1. If we show their mass distributions together in one plot, as in Fig.~\ref{fig:LIGO}, one can see that they are not randomly distributed, they tend to cluster around 25 $\Msun$, specially if we ignore the final black hole after merger. It is true, however, that the lower and upper edges of the distribution may be affected by the AdvLIGO sensitivity to BHB mergers with masses between a few to 150 solar masses, due to their seismic and shot noise experimental constraints, respectively~\cite{TheLIGOScientific:2016pea}.

\begin{table}
\centering\renewcommand\arraystretch{1.2}
\begin{tabular}{|l|c|c|c|c|c|c|}
\hline
Event & $m_1\ (\Msun)$ & $m_2\ (\Msun)$ & $M\ (\Msun)$ \\ \hline
GW150914 & $36^{+5}_{-4}$ & $29^{+4}_{-4}$ & $62^{+4}_{-4}$ \\ \hline
LVT151012 & $23^{+18}_{-6}$ & $13^{+4}_{-5}$ & $35^{+14}_{-4}$ \\ \hline
GW151226 & $14.2^{+8.3}_{-3.7}$ & $7.5^{+2.3}_{-2.3}$ & $20.8^{+6.1}_{-1.7}$ \\ \hline
GW170104 & $31.2^{+8.4}_{-6.0}$ & $19.4^{+5.3}_{-5.9}$ & $48.7^{+5.7}_{-4.6}$ \\ \hline
GW170814 & $30.5^{+5.7}_{-3.0}$ & $25.3^{+2.8}_{-4.2}$ & $53.2^{+3.2}_{-2.5}$ \\ \hline
\end{tabular}
\caption{\label{tab:LIGO} The masses of the components of the five black hole binaries detected by LIGO, together with the final mass of the black hole after merger.}
\end{table}%

We will then assume, for simplicity, that the PBH mass distribution is lognormal\footnote{A different convention is to replace $\ln$ by $\log_{10}$ in Eq.~\ref{eq:LN}, which is equivalent to a rescaling of $\sigma$ into $ \ln 10 ~\sigma $.} with parameters ($\mu,\,\sigma$),
\be\label{eq:LN}
P(M) = \frac{\rr d \,n_{\rm PBH}}{\rr d\ln M} = \frac{\fpbh}{\sqrt{2\pi}\,\sigma}\,\exp\left[-\frac{\ln^2(M/\mu)}{2\sigma^2}\right]\,,
\ee
where $\fpbh = \OPBH/\OCDM$ is the {\em total} fraction of PBH in cold dark matter, and we have chosen $P(M)$ to be normalized to that fraction, $\int_0^\infty P(M) \,dM/M = \fpbh$. Note that the mean mass for this distribution is given by 
\be\label{eq:mean}
\bar M = \fpbh^{-1}\,\int_0^\infty \frac{dM}{M}\,P(M)\,M  = \mu\,e^{\frac{1}{2}\sigma^2}\,,
\ee
which can be significantly larger than $\mu$.

\begin{figure}[!ht]
\centering
\includegraphics[width = 0.48\textwidth]{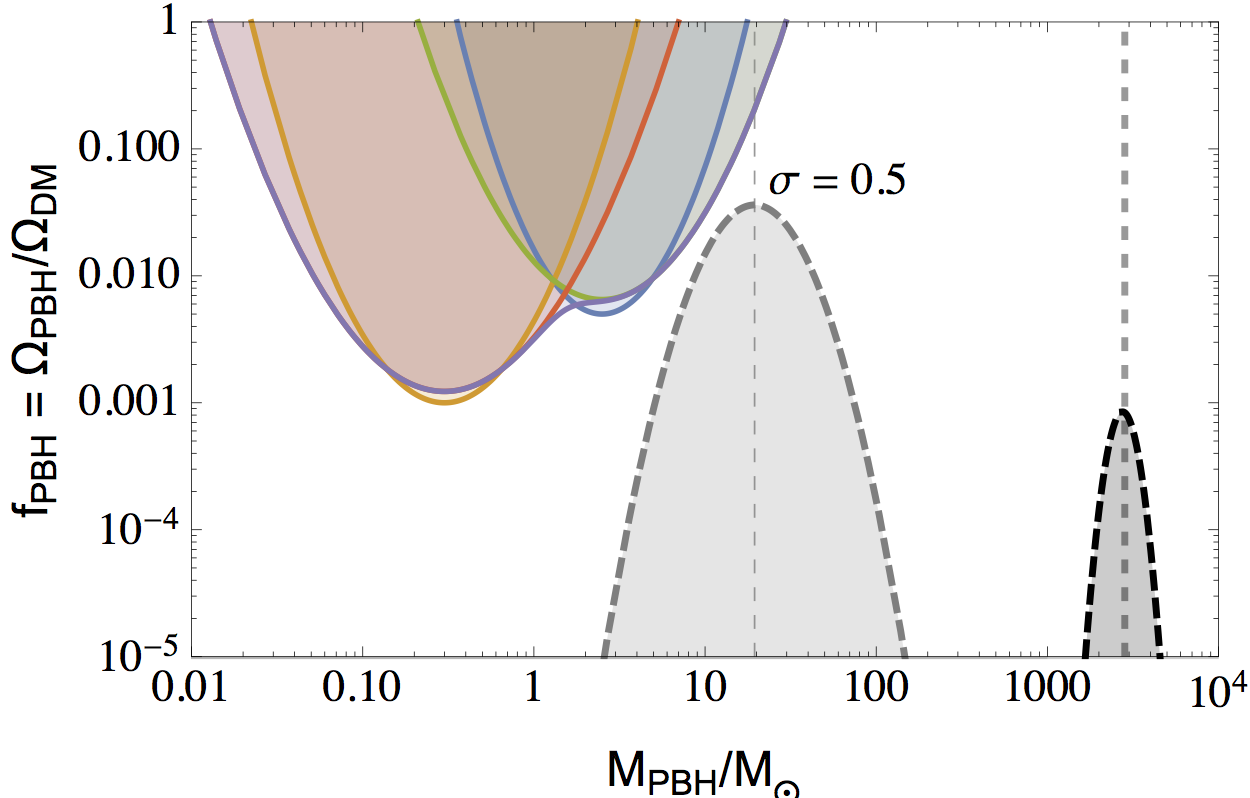}
\caption{Synthetic microlensing constraints for an extended lognormal mass distribution of width $\sigma=0.5$, assuming PBH spatial uniformity, the x-axis representing the central mass $\mu$. The blue and orange contours correspond to the expected constraints for a {\em monochromatic} distribution ($\sigma \simeq 0$). The green and red lines are the {\em modified} constraints when taking into account the width of the PBH distribution~\cite{Carr:2017jsz}. The purple line is the overall microlensing constraint.  The (light grey) curve shows the distribution for $\mu = 25 M_\odot$ and $\sigma=0.5$, excluded by microlensing if $f_{\rm PBH} = 1$ and PBH have a uniform spatial distribution.  The narrow (black) distribution on the right is the equivalent distribution for the same extended-mass model but clustered PBH with $\Ncl =100$ members per cluster. It is a lognormal with $\bar\mu = 2200\,\Msun$ and $\bar\sigma=0.053$ which passes all microlensing constraints. 
}
\vspace*{-1mm}
\label{fig:extcons}
\end{figure}

Let us consider now a potential constraint from a specific microlensing experiment. It is typically presented as a bound on the fraction of PBH in a given infinitesimal interval ($M,\, M + \rr d M$) around mass $M$, i.e. the bounds are shown as constraints $C(M)$ on a {\em monochromatic} mass distribution. Those coming from microlensing experiments ($i=1,\dots,\,N$) are typically of the form
\be\label{eq:cons}
C_i(M) = A_i\,\exp\left(\frac{\ln^2(M/m_i)}{2s_i^2}\right)\,,
\ee
where ($A_i,\,m_i,\,s_i$) are the amplitude, central mass and width parameters that characterize the constraint. Note, however, that most PBH scenarios have a wide mass distribution~\cite{Clesse:2015wea,Garcia-Bellido:2016dkw,Bartolo:2016ami}, rather than monochromatic, and therefore the actual constraint can be written as~\cite{Carr:2017jsz}
\be
\int_0^\infty \frac{dM}{M}\,\frac{P(M)}{C(M)} \leq 1\,.
\ee
For a lognormal distribution of PBH, see Eq.~(\ref{eq:LN}), the individual ($i=1,\dots,N$) integral constraint becomes
\be
\frac{\fpbh\ s_i}{A_i\sqrt{s_i^2+\sigma^2}}\,\exp\left(-\frac{\ln^2(\mu/m_i)}
{2(s_i^2+\sigma^2)}\right) \leq 1\,.
\ee
In the case of multiple constraints one finds
\be
\fpbh(M) \leq \left[\sum_{i=1}^N
\frac{s_i^2}{A_i^2(s_i^2+\sigma^2)}\,
\exp\left(-\frac{\ln^2(M/m_i)}{(s_i^2+\sigma^2)}\right)
\right]^{-1/2}\,,
\ee
where we have assumed that each constraint is statistically independent and we have summed them in quadrature. We have plotted in Fig.~\ref{fig:extcons} the enhanced constraints for the case of a wide lognormal distribution with $\sigma=0.5$. Note that the PBH model in the example (light gray contour) would then be barely acceptable, and future improvements on long duration microlensing experiments (colored curves on the left) would be able to rule out a large fraction of that PBH mass distribution.

\subsection{Clustering of PBH}

While a reanalysis of the effect of extended mass distributions on the microlensing constraints has been done recently~\cite{Green:2016xgy,Carr:2017jsz}, there is no consideration yet of the fact that those PBH distributions not only cover a wide mass range, but also that PBH are spatially clustered. The origin of PBH from high and broad peaks in the primordial power spectrum suggest that PBH are highly clustered in space~\cite{Chisholm:2005vm}, as successive fluctuations reenter the horizon during the radiation era. Such clusters of PBH will soon rearrange themselves via dynamical friction, with the most massive PBH at the center and a sub-halo of PBH orbiting around it with decreasing masses as we move away from the center of the cluster. Let us now estimate the typical number of PBH and their separation in each cluster.

First of all, we estimate the mass fraction of Hubble domains that collapse to form black holes at formation as~\cite{Clesse:2015wea,Chisholm:2005vm}
\be\label{eq:beta}
\beta(\nu) = {\rm erfc}\Big(\nu/\sqrt2\Big)\simeq \sqrt{\frac{2}{\pi}}\frac{e^{-\nu^2/2}}{\nu}\,,
\ee
for $\nu\equiv\delta_c/\sigma_H = \delta_c/\sqrt{P(k_H)} \gg 1$, with $P(k_H)$ the power spectrum on the scale of the horizon at the time of PBH formation. From concrete models of inflation~\cite{Garcia-Bellido:2017fdg} one obtains $\nu\simeq6$.
The typical distance between PBH at the redshift of formation $z_f$ is then
\be
\lambda(z_f) = \frac{d_H(z_f)}{\beta(z_f)^{1/3}} \sim 
1.2\times10^5\ {\rm km}\,\left(\frac{6\times10^{11}}{1+z_f}\right)^{5/3}, 
\ee
where we have estimated
\be
\beta(z_f)  \sim 3\times 10^{-9} \,
\left(\frac{6\times10^{11}}{1+z_f}\right)\,,
\ee
and the horizon distance at formation as
\be
d_H(z_f) = d_H(z_{\rm eq}) \left(\frac{1+z_{\rm eq}}{1+z_f}\right)^2 \sim
240\,{\rm km}\,\left(\frac{6\times10^{11}}{1+z_f}\right)^2\,,
\ee
where $z_{\rm eq} $ is the redshift of matter-radiation equality, for typical masses (of order the horizon mass)
\be
\Mpbh \sim 20\,\Msun\,\left(\frac{6\times10^{11}}{1+z_f}\right)^2\,.
\ee 

The number of PBH in each cluster was computed in Ref.~\cite{Chisholm:2005vm} and found to be large at formation
\be
N_{\rm cl} = \frac{10}{7}\,\beta(\nu)\,e^{3\nu^2/4} \propto \beta(z_f)^{-1/2}\,,
\ee
which is, in typical models, of order $N_{\rm cl}\sim2000$. This number will change with the evolution of the universe, as dynamical friction and internal ``heating" will increase the size of the cluster and some fraction of the black holes will evaporate from the cluster via three body interactions~\cite{Sigurdsson:1993zrm} or be sling-shot away by a heavier PBH.

The local density contrast can also be estimated~\cite{Chisholm:2005vm} as
\be
\delta \sim e^{\nu^2/2} \propto \beta(z_f)^{-1} \sim 6\times10^7\,,
\ee
which is expected to increase since formation, via PBH merging and gas accretion. Part of the growth in density contrast will correspond to an increase in the mass of individual black holes, via Bondi accretion, but most of the increase is due to a decrease in volume. Such massive clusters may subtend today scales well below the parsec, below than their typical Einstein radius relevant for microlensing events in the Magellanic clouds, and are separated from each other by distances of tens of parsecs, in the outer halos of galaxies, e.g. like in the vicinity of the sun, see~\cite{Garcia-Bellido:2017fdg}
\be
d_{\rm PBH} = 50\,{\rm pc}\,\left(\frac{\Mtot}{100\,\Msun}\right)^{1/3}\,,
\ee
where $\Mtot$ is the total mass of the cluster. Of course, these clusters can reach much smaller inter-separation distances in denser areas, like dwarf spheroidals and molecular clouds, where they are highly concentrated~\cite{Clesse:2016vqa}.

These clusters of PBH are then detectable via microlensing events. Given the typical distance between them and their point-like character, with a hierarchical distribution of masses within the cluster, one expects much fewer microlensing events towards stars in the Large Magellanic Cloud than would be expected form a uniform distribution and, moreover, we expect that some of the less-massive PBH orbiting inside the cluster may induce caustics in the stellar light curves. Such signatures would be extremely suggestive of PBH clustering.

What is still unclear is the number of black holes remaining in a typical cluster today after merging and accretion, and the fraction of PBH that has been ejected and is uniformly distributed in the galactic halos.  These are expected the lightest ones and could explain the microlensing events detected in M31 and quasars~\cite{Mediavilla:2017bok}, which suggest that between $15\%$ and $30\%$ of the halo mass of galaxies could be made of compact objects with sub-stellar masses.  Those issues depend very much on the merger history of PBH clusters and they detailed dynamics, whose study requires $N$-body simulations on scales that are well below the typical resolution of usual DM simulations. 

Here we will rederive the microlensing constraints on the PBH mass distribution assuming a fixed number $\Ncl$ of primordial black holes per cluster. The total mass of the PBH cluster will vary from cluster to cluster, following a lognormal distribution, but we can estimate the typical total mass of an average cluster as
\be\label{eq:mtot}
\Mtot = \sum_{i=1}^{\Ncl} M_i = \Ncl\,\bar M\,,
\ee
where $\bar M$ is the mean of the lognormal given by Eq.(\ref{eq:mean}). Any given cluster will have
a total mass which is different from Eq.~(\ref{eq:mtot}) and in fact its value will be distributed again like a lognormal. We have made multiple realizations, drawing masses from a lognormal distribution, with different values of $\Ncl$, and have confirmed that the total mass of Eq.~(\ref{eq:mtot}) is distributed as a lognormal with mean and dispersion 
\be\label{eq:remean}
\bar\mu = \Ncl\,\bar M\,, \hspace{1cm}
\bar\sigma = \sqrt{\frac{e^{\sigma^2}-1}{\Ncl}}\,, 
\ee
valid for any value of $\Ncl$. Thus the distribution of values of $\Mtot$ is a narrow lognormal peaked around $N_{\rm cl} \bar M$, as can be seen in Fig.~\ref{fig:extcons} for a concrete case.

Since the PBH cluster is essentially point-like for microlensing purposes, the mass responsible for the long-duration events is the total cluster mass. Therefore, the microlensing constraints do not apply to the individual masses of PBH within the cluster, but to the total mass of the cluster. Assuming that these PBH clusters are uniformly distributed over the halo of our galaxy, we can derive new constraints on PBH. 
In reality, the width of the cluster mass distribution will also depend on the physics of the environment, whether those PBH are in a dense medium with gas, like inside molecular clouds, which induces friction and will tend to merge the PBH into a single massive IMBH, or populate the outskirts of the galactic halo, where individual PBH would not find each other and merge within the age of the universe.  Knowledge of such a cluster distribution requires detailed numerical simulations.  We therefore restrict for the moment our analysis to the simplest case of an {\em equivalent} narrow lognormal distribution with mean and dispersion given by Eq.~(\ref{eq:remean}).

In this case, the constraints from microlensing on the extended PBH mass distribution can be recomputed as if all the PBH distribution would be in a narrow spectrum around a mass $\Mtot \gg \mu$, as represented on Fig.~\ref{fig:extcons}.

\subsection{CMB constraints on clustered PBH}

Note that the CMB constraints are not directly affected by PBH clustering since the effect of energy injection occurs near the individual black holes and it is their individual masses that count.  However, the effective PBH velocity with respect to baryons is expected to be enhanced if PBH are clustered, with respect to the uniform case.  This indirectly impacts the CMB constraints since they are very sensitive to the PBH velocity.   The overall CMB constraints on the total fraction of PBH to CDM can be written as~\cite{Poulin:2017bwe}
\be
\fpbh < \left(\frac{4\Msun}{\Mpbh}\right)^{1.6}
\left(\frac{v_{\rm eff}}{10\,{\rm km/s}}\right)^{4.8}\left(\frac{0.01}{\lambda}\right)^{1.6}\,,
\ee
where $\lambda\sim0.01$ is the Bondi accretion rate in Eddington units.  For PBH with $\mu \simeq 10 \Msun$ and clusters with $N_{\rm cl}$ from 100 to 1000 PBH with typical cluster masses $\Mtot\sim{\rm few}\,10^3\,\Msun$ and sizes $r_h  \sim 1$ mpc, the Viral velocity 
is of order $v_{\rm eff} \sim 20$ to $70$ km/s and thus CMB bounds are satisfied without any problem, whereas for PBH uniformly distributed one expects lower $v_{\rm eff}$ at high redshift and the model would be excluded. 

\begin{figure}[!ht]
\centering
\includegraphics[width = 0.48\textwidth]{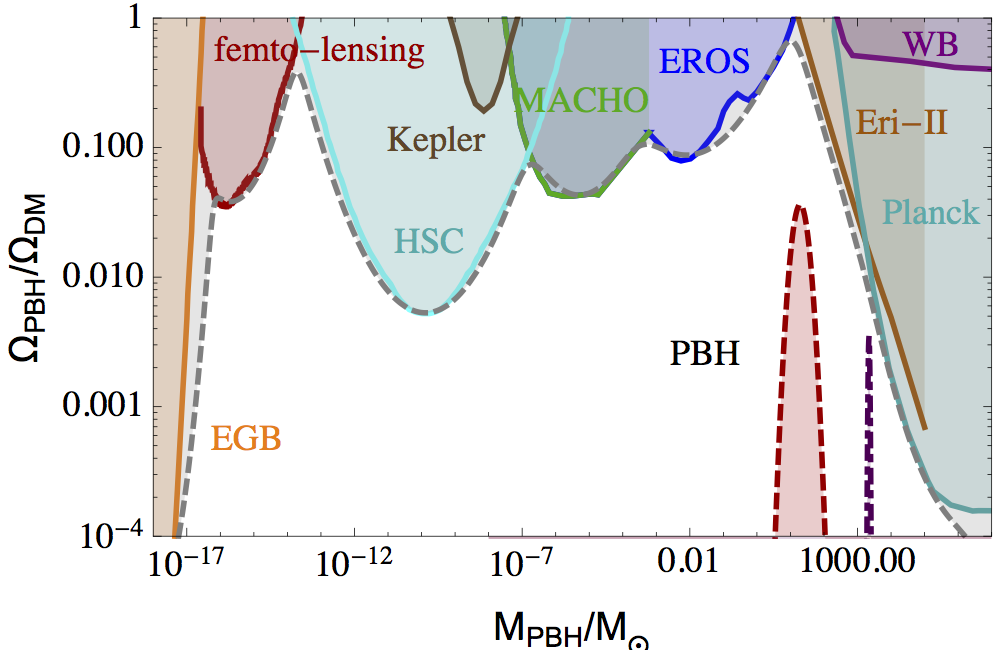}
\caption{Observational constraints on PBH from a plethora of experiments (for a review see~\cite{Carr:2016drx}). We have taken into account the fact that our mass distribution is wide, which changes the constraints (grey line), for details see Ref.~\cite{Carr:2017jsz}. We have assumed PBH to be distributed as a lognormal distribution with $\mu\sim25\,\Msun$ and $\sigma\sim0.5$ (red dashed curve). The mass-equivalent clustered distribution has $\bar\mu\sim2200\,\Msun$ and $\bar\sigma\sim0.053$ (purple dot-dashed curve). Clearly the final PBH distribution passes all constraints, and at the same time constitute all of dark matter.}
\vspace*{-1mm}
\label{fig:constraints}
\end{figure}

\section{Conclusions}

There exists several clues (rate, mass and spin of BH mergers, spatial correlations in CIB and X-ray backgrounds) supporting a scenario where all the Dark Matter is made of stellar-mass PBH.  This scenario is however in tension with microlensing and CMB constraints.  They have been recently reevaluated for a broad PBH mass distribution, but assuming spatial uniformity.  If instead PBH are clustered, as expected if they formed from broad peaks in the primordial power spectrum, we find that the microlensing bounds can be relaxed significantly and even evaded for relatively broad-mass distributions, if there are typically more than hundreds of PBH per cluster.   On the other hand, the PBH velocity is enhanced compared to the uniform case and therefore the scenario evades the latest constraints from CMB anisotropies.   This way, a new window around 1 to 10 solar masses has opened up for PBH to comprise the totality of the dark matter.

\section*{Acknowledgements}

The authors thank Ester Ruiz Morales for interesting discussions.
JGB thanks the CERN TH-Division for hospitality during his sabbatical, and acknowledges support from the Research Project FPA2015-68048-03-3P [MINECO-FEDER] and the Centro de Excelencia Severo Ochoa Program SEV-2012-0597. He also acknowledges support from the Salvador de Madariaga Program, Ref. PRX17/00056.   The work of SC is supported by a \textit{Charg\'e de Recherche} grant of the Belgian Fund for Research FRS/FNRS.

\end{document}